\documentclass[twocolumn,twoside,slac_Two]{revtex4}
\usepackage{graphicx}
\usepackage{fancyhdr}
\usepackage{graphics}
\usepackage{epstopdf}
\usepackage{textpos}
\begin{document}

\title{\centering A Search for \boldmath{$t\bar{t}$} Resonances in the Dilepton Channel in 1.04 fb$^{-1}$ of \boldmath{$pp$} collisions at $\sqrt{s} = 7$ TeV with the ATLAS experiment}


\author{
\centering
\begin{center}
M. Petteni on behalf of the ATLAS Collaboration
\end{center}}
\affiliation{\centering Simon Fraser University, BC, V5A 1S6, Canada}
\begin{abstract}
A search for a high mass $t\bar{t}$ resonance in $pp$ collisions at $\sqrt{s} = 7$~TeV at the LHC is presented. 
We search for such a resonance using the final state where the $W$ bosons from the top quark decay into either an 
electron or a muon (dilepton final state). 
The data were recorded by the ATLAS experiment during 2011 and correspond to a total integrated luminosity of ${\cal L} = 1.04~fb^{-1}$.
No statistically significant excess above the Standard Model expectation is observed. 
Upper limits at the 95\% Confidence Level (C.L.) are set on the cross section times branching ratio of the resonance decaying to 
$t\bar{t}$ pairs as a function of the resonance mass. A lower mass limit of 0.84~TeV is set for the case of a Kaluza-Klein 
gluon resonance in the Randall-Sundrum Model.
\end{abstract}

\maketitle
\thispagestyle{fancy}


\section{Introduction}
Many models of physics beyond the Standard Model (BSM) predict the existence of new resonances that decay predominantly into top quark pairs.
The top quark is unique among the known matter constituents. It is the only fermion 
whose mass is very close to the scale of electroweak symmetry breaking. Partly due to this reason,
the top quark has a special treatment in many BSM scenarios.
 These include Kaluza-Klein (KK) excitations of the graviton, gluon as well as 
other gauge bosons which couple to top quarks \cite{EDsigs}\cite{RS}.

This note provides a description of the search for new heavy particles decaying 
to $t\bar{t}$ pairs in the dilepton channel by the ATLAS experiment \cite{ATLAS}. 

Two variables of particular interest in this analysis are the total transverse momentum, $H_{\mathrm{T}}$, and $E_{\mathrm{T}}^{miss}$. $H_{\mathrm{T}}$ is defined as the scalar
sum of the transverse momenta of the two identified leptons and all the jets in the event
above a 25$~GeV$ threshold. $E_{\mathrm{T}}^{miss}$ is the missing transverse momentum from the escaping neutrinos from the leptonic $W$ boson decay.

In the absence of any significant signal, limits on the production cross section times branching ratio ($\sigma B$) are 
set for a series of resonance masses using a template shape fitting method. 
The limits on $\sigma B$~are translated into limits on the resonance mass using predictions for a Kaluza-Klein 
gluon resonance in the Randall-Sundrum Model.

\section{Event Selection}

The data analyzed were collected in the period from March to July 2011, corresponding to a total integrated luminosity of ${\cal L} = 1.04~fb^{-1}$.
The selection of $t\bar{t}$ events makes use of reconstructed
electrons, muons, jets and $E_{\mathrm{T}}^{miss}$. Electron candidates are required to have transverse energy $E_{\mathrm{T}} > 25$~GeV\ and pseudorapidity $|\eta| < 2.47$ (the pseudorapidity is defined in terms of the polar angle $\theta$ as $\eta=-\ln\tan(\theta/2)$). Muon candidates are required to have a transverse momentum larger than 20$~GeV$ with $|\eta| < 2.5$ and to pass cosmic-muon rejection. Jet candidates are required to have a $p_{\mathrm{T}}>25$~GeV and $ |\eta| < 2.5$. 

The event selection is driven by the topology of the top decay. Each selected event is required to have:
\begin{itemize}
\item Satisfied good data quality requirements.
\item Been triggered by either a single-electron trigger with a
threshold of $E_{\mathrm{T}}> 20$~GeV~or a single-muon trigger with a threshold of $p_{\mathrm{T}} > 18$~GeV. 
\item An offline-reconstructed primary vertex with at least five tracks. 
\item Two or more selected jets
\item Exactly two oppositely charged leptons (electrons or muons) with a dilepton invariant mass greater than 10 GeV.
\end{itemize}

Following this, two samples are defined, a control region, depleted of signal and dominated by Z+jets events, and a signal region. 

The control region is defined by a cut on the invariant mass of the two leptons, $|m_{Z}-m_{\ell\ell}| < 10$ GeV and is only used for the ee and $\mu\mu$ channels. The signal region has two separate definitions. For the ee and $\mu\mu$ final states the following cuts are used: $E_{T}^{miss} > 40$ GeV and $|m_{Z}-m_{\ell\ell}| > 10$ GeV. For the e-$\mu$ channel $H_{\mathrm{T}}> 130$ GeV is required. The difference arises from the additional source of $E_{T}^{miss}$ in the background processes for the e$\mu$ channel from $Z/\gamma^{*} \rightarrow \tau\tau$ and diboson decays. Full details of the event selection can be found in Ref. \cite{ttdilepconf}.

\section{Backgrounds}

\begin{figure*}[!htbp]
\centering
\includegraphics[width=0.48\textwidth]{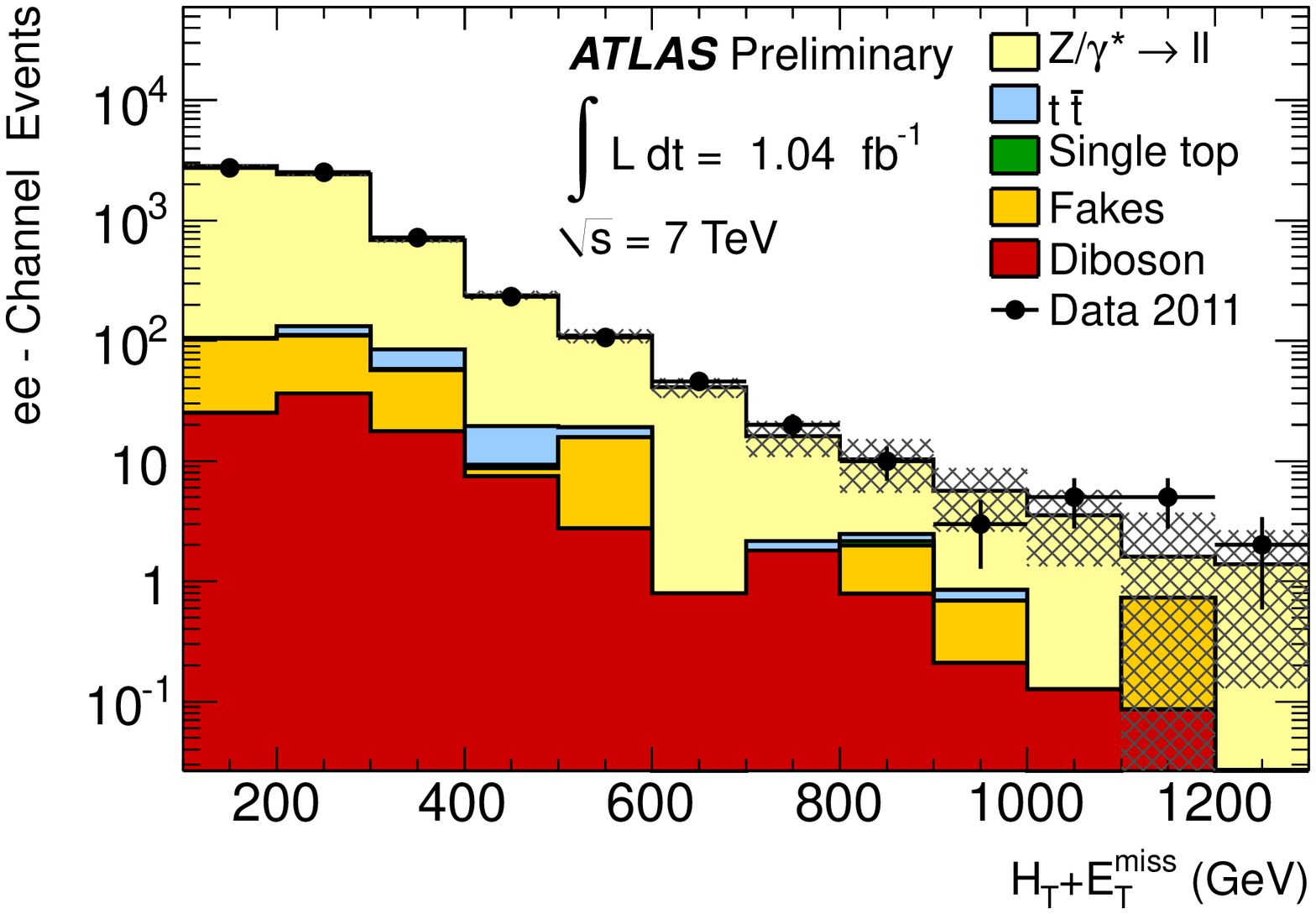}
\includegraphics[width=0.48\textwidth]{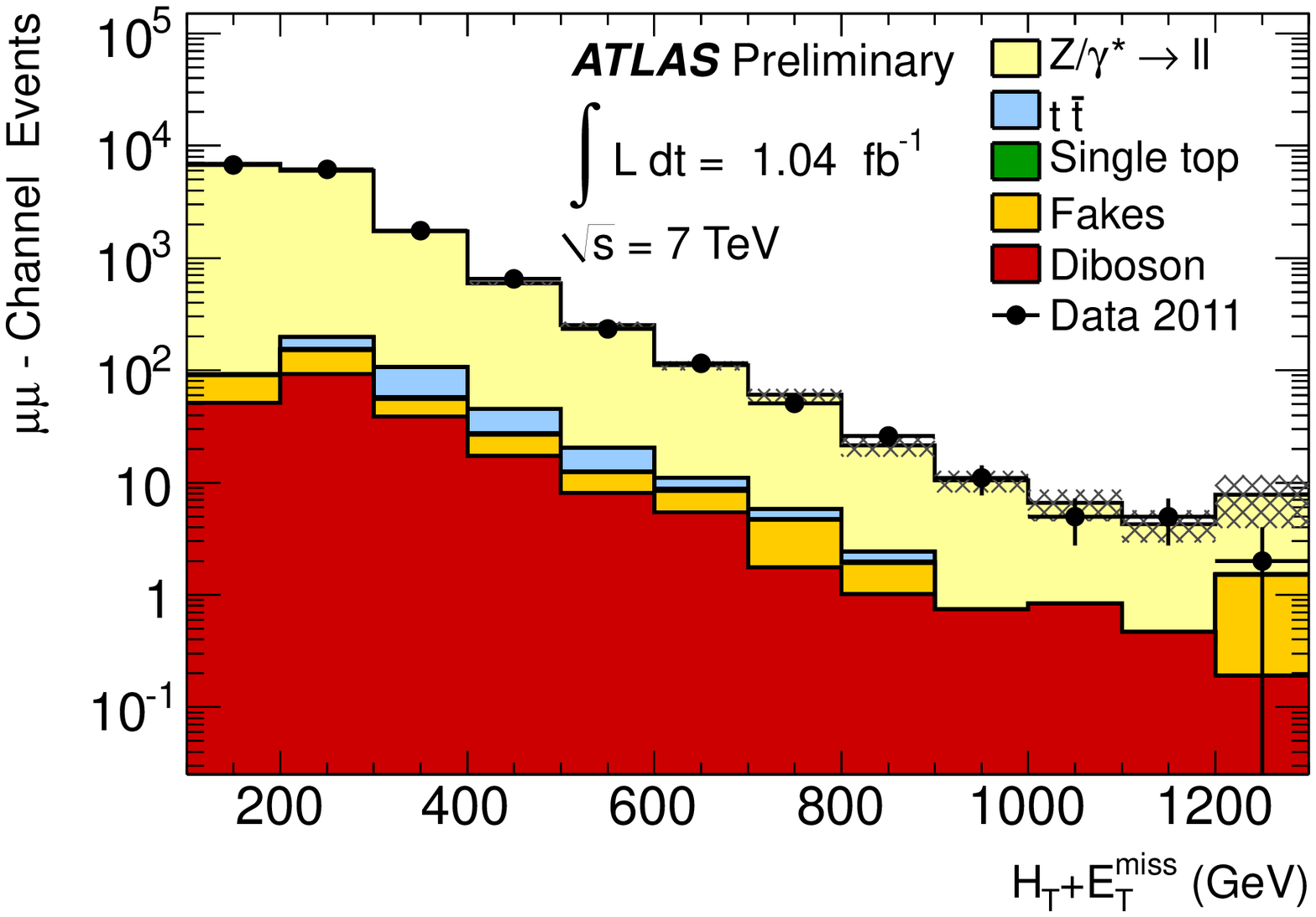}
\caption{The $H_{\mathrm{T}} + E_{\mathrm{T}}^{miss}$ distributions (left) for the dielectron and (right) for the dimuon control region. The control region is defined by inverting the $Z$ boson
rejection window cut and predominantly selects $Z$ + jets events.   
The points represent ATLAS data and the filled histograms show the simulated backgrounds including the statistical uncertainty 
represented by the hashed band. The last bin displays the overflow bin \cite{ttdilepconf}.
}
\label{fig:control1}
\end{figure*}

The main background to KK-gluon production is the irreducible Standard Model $t\bar{t}$ production, which features an indistinguishable final state topology. Background sources also include: $Z$ + jets, single top production, diboson production and "fake" i.e. QCD multijet and $W$+jets processes where one or two jets are misidentified as leptons.

All background distributions are derived from Monte Carlo (MC), except for the fakes which are obtained from data. The contribution of these events is estimated from data \cite{dilxsec}. The signal MC is generated using MadGraph interfaced to Pythia.
The number of jets in the $Z$+jets background is normalized to data in a control region and its contribution in the signal region is normalized to data (for ee and $\mu\mu$ decays) using a Matrix Method technique, inclusive in jet number~\cite{ttdilepconf}. The background predictions in the signal region are listed in Table \ref{table:bkg}.
The MC is normalized using the ATLAS luminosity measurement. The validation of the $H_{\mathrm{T}} + E_{\mathrm{T}}^{miss}$ distribution in the data control sample is shown in Figure~\ref{fig:control1}. 

\begin{table}[h!]
    \centering
    \caption{: Background composition in the signal region.
    Both statistical and systematic uncertainties are included \cite{ttdilepconf}.}
    \begin{tabular}{|l|c|}
      \hline
      Process &  Predicted number of background events \\ 
      \hline
      $t\bar{t}$                  & 1920$^{+230}_{-220}$ \\
      $Z/\gamma^{*} \rightarrow ee$ + jets         & 130$^{+72}_{-49}$ \\ 
      $Z/\gamma^{*} \rightarrow \mu\mu$ + jets     & 140$^{+27}_{-21}$ \\           
      $Z/\gamma^{*} \rightarrow \tau\tau$ + jets   & 85$^{+12}_{-10}$ \\ 
      \rm{Diboson}                 			& 83$^{+13}_{-12}$\\ 
      \rm{Single top}                 		& 98$^{+14}_{-13}$  \\ 
      \rm{Fakes}            				& 96$^{+94}_{-51}$ \\ 
      \hline 
      \hline 
      Total background               &   2550$^{+330}_{-300}$  \\ 
      Data            &    2659 \\ 
      \hline 
    \end{tabular} 
    \label{table:bkg} 
\end{table} 

\section{Systematics}

Sources of systematic uncertainties were studied and a non-negligible impact on the analysis was found from the uncertainty on:

\begin{itemize}
\item Lepton energy scale, reconstruction, identification (ID) and trigger efficiency.
\item Jet energy scale and resolution.
\item ISR/FSR modeling.
\item Choice of PDF.
\item Dependence on choice of MC generator and parton showering model.
\end{itemize}

A 50\% systematic uncertainty was also assigned to the  
fake background estimate derived from the uncertainty on the misidentification rates. The effect of this systematic uncertainty was found to be negligible.

The largest systematic variations on the background prediction was from the jet energy scale (~7\%), lepton ID/trigger efficiency (~4\%) and the $t\bar{t}$ generator dependence (4\%). For the signal the greatest was the jet energy resolution (~6\%)

For the above systematic uncertainties both rate and shape dependences were computed. Further to these the luminosity (3.7\%) and cross-section uncertainties for $t\bar{t}$ were included as overall rate changes.

\section{Statistical Analysis}

\begin{figure}[h]
  \centering
  \includegraphics[width=0.5\textwidth]{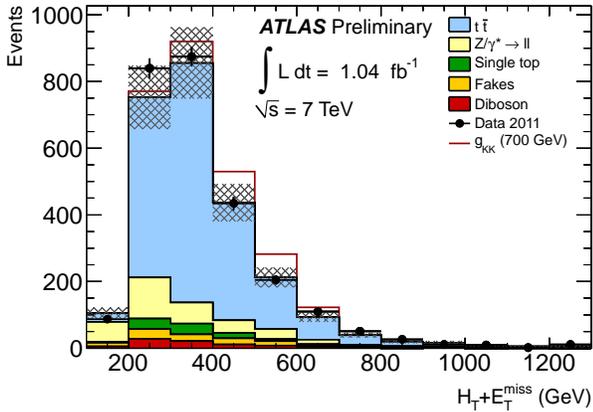}
  \caption{Data - MC comparison for the $H_{\mathrm{T}} + E_{\mathrm{T}}^{miss}$ distribution together with a KK-gluon signal with a mass of 700 GeV~. The statistical and systematic uncertainty on the Monte Carlo is represented by the hashed band \cite{ttdilepconf}.}
  \label{fig:htscan}
\end{figure}

We choose $H_{\mathrm{T}} + E_{\mathrm{T}}^{miss}$ as the discriminating variable for this analysis. This variable is strongly correleted to the $t\bar{t}$ invaraiant mass, which due to the missing neutrinos cannot be directly reconstructed. A comparison of the $H_{\mathrm{T}} + E_{\mathrm{T}}^{miss}$ MC prediction and observed data including a hypothetical signal is shown in Figure \ref{fig:htscan}. 

For the statistical analysis we construct a binned likelihood function based on Poisson statistics for signal and background for each bin of the $H_{\mathrm{T}} + E_{\mathrm{T}}^{miss}$ distribution. Sources of systematic uncertainty are included as nuisance parameters in the likelihood function. 

A p-value based on a likelihood ratio technique that compares the SM only hypothesis to an extension of the standard model that includes a KK-gluon is calculated. The p-value was found to be 0.4, consistent with the SM only hypothesis.

\begin{figure}[t]
  \centering
  \includegraphics[width=0.5\textwidth]{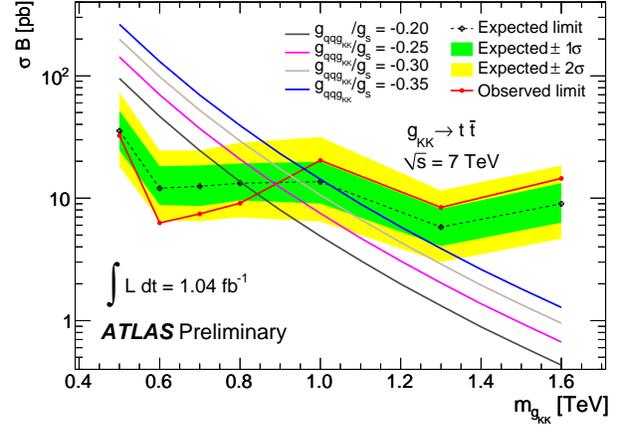}
  \caption{Expected and observed limits on cross section times branching ratio at 95\% C.L. and expected cross section for a Randall-Sundrum KK-gluon $g_{KK}$. Cross
  sections were calculated using the MRST 2007 LO$^*$ PDF~\cite{ttdilepconf}. }
  \label{fig:limit}
\end{figure}

Figure \ref{fig:limit} shows the 95\% C.L. exclusion limit on the cross section times branching ratio. Details of the cross section calculation can be found in Ref. \cite{ttdilepconf}. The coupling of light quarks to the KK-gluon is varied by scaling the strong coupling parameter $g_{qqg_{KK}}/g_{s}$ 
in a range from 0.2 to 0.35, corresponding to a theoretically viable range of values. Table~\ref{tab:massLimits} lists the expected and observed KK-gluon mass limits obtained for each model point.

\begin{table}[h]
  \centering
  \caption{Expected and observed lower limits on the KK-gluon mass in the Randall-Sundrum model}
  \begin{tabular}{l|c|c|}
    \cline{2-3} 
                        & \multicolumn{2}{|c|}{Mass Limit (TeV)}\\
    \hline
    \multicolumn{1}{|c|}{$g_{qqg_{KK}}/g_{s}$}  & Expected & Observed\\
    \hline
    \multicolumn{1}{|c|}{-0.20}         & 0.80 & 0.84  \\
    \multicolumn{1}{|c|}{-0.25}         & 0.88 & 0.88 \\
    \multicolumn{1}{|c|}{-0.30}         & 0.95 & 0.92 \\
    \multicolumn{1}{|c|}{-0.35}         & 1.02 & 0.96  \\
    \hline
  \end{tabular}
  \label{tab:massLimits}
\end{table}

\section{Conclusions}

The ATLAS detector has been used to search for high-mass resonances in the dilepton $t\bar{t}$ final state. The $H_{\mathrm{T}} + E_{\mathrm{T}}^{miss}$  
observable is well described by the Standard Model backgrounds. We find no significant excess at high $H_{\mathrm{T}} + E_{\mathrm{T}}^{miss}$ in the data, and set limits 
on the cross section times branching ratio for KK-gluon production as well as upper limits at 95\% C.L. on the mass of the KK-gluon in the Randall-Sundrum model of 0.84 TeV. 





\bigskip 
\bibliography{basename of .bib file}

\end{document}